\documentclass[]{tCPH2e}

\newcommand{\kms}          {\mbox{${\rm km~s^{-1}}$}}

\newcommand{\ee}           {\mbox{$^{-2}$}}
\newcommand{\eee}          {\mbox{$^{-3}$}}
\def\H2{\mbox{${\rm H}_2$}}
\def\mH2{\mbox{$m_{\rm H_2}$}}
\def\nH2{\mbox{$n_{\rm H_2}$}}
\def\NH2{\mbox{$N_{{\rm H}_2}$}}
\def\MH2{\mbox{$M_{{\rm H}_2}$}}
\def\HII{\mbox{H\,{\sc II}}}
\def\Msun{\mbox{$M_\odot$}}

\def\micron{\hbox{$\mu$m}}
\def\simgt{\lower.5ex\hbox{$\; \buildrel > \over \sim \;$}}
\def\simlt{\lower.5ex\hbox{$\; \buildrel < \over \sim \;$}}

\def\fe60{\mbox{$^{60}$Fe}}
\def\al26{\mbox{$^{26}$Al}}

\begin{document}
\doi{10.1080/00107511003764725}
 \issn{1366-5812}
\issnp{0010-7514}

\jvol{51} \jnum{5} \jyear{2010} \jmonth{Dec}

\markboth{J. P. Williams}{Contemporary Physics}

\articletype{}

\title{The Astrophysical Environment of the Solar Birthplace}

\author{Jonathan P. Williams$^{\ast}$\thanks{$^\ast$Corresponding author. Email: jpw@ifa.hawaii.edu}\\
\em{Institute for Astronomy, University of Hawaii, Honolulu , USA}}

\maketitle

\begin{abstract}
Our Sun, like all stars, formed within a cold molecular cloud.
Astronomical observations and theory provide considerable
detail into this process. Yet cosmochemical observations of short
lived radionuclides in primitive meteorites, in particular \fe60,
provide unequivocal evidence that the early solar system inherited
fresh nucleosynthetic material from the core of a hot, massive star,
almost certainly ejected in a supernova explosion.
I give a short introduction to the fields of star formation and
meteoritics and discuss how the reconciliation of their disparate clues
to our origin places strong constraints on the environment of the
Solar birthplace.
Direct injection of supernova ejecta into a protoplanetary
disk or a dense molecular core is unlikely since their small sizes
require placement unusually close to the massive star.
Lower density molecular cloud clumps can capture more ejecta
but the radionuclides decay during the slow gravitational collapse.
The most likely scenario is on the largest scales via the formation
of enriched molecular clouds at the intersection of colliding supernova
bubbles in spiral arms.

\bigskip
\end{abstract}

\section{Introduction}
\label{sec:intro}
We know a great deal about the Sun.
We know, to astonishing precision, its mass, size, composition, and age.
Each of these measurements was a triumph of astrophysics.
Nobel prize winning work has detailed the nuclear reactions in its
core and its eventual fate as a white dwarf.
Yet the {\it origins} of our Sun and of the solar system, are far
less well understood. This is an active area of current research
and, as with most areas of astronomy, one that is driven by
technology and has exciting prospects ahead.
Moreover, with the ever-growing catalog of extrasolar planets,
and now direct imaging of multiple planet systems,
the question of the characteristics of the solar system
relative to others can begin to be addressed.

The Sun is only about one third the age of the Universe.
We can observe distant galaxies that formed before our Sun existed.
We can also observe nearby protostars as they condense from
a diffuse cloud.
By studying today's protostars and the disks around them,
we can learn about
the processes that led to the formation of our Sun and solar system.
The underlying theoretical principles are described
in \cite{2007ARA&A..45..565M,2005ConPh..46...29C}.
I outline the scales of mass, size, and time for the different stages
of star formation in \S\ref{sec:starform}.

We can also learn about the conditions of the early solar system
by studying meteorites, the rocky remnants of the protosolar disk.
The age of the solar system was deduced from radioactive isotopic
analysis of the most primitive meteorites \cite{1997ConPh..38..103H}.
Cosmochemical studies of meteoritic mineralogy and composition reveal
the way in which interstellar dust grains metamorphised
to larger and larger pieces and, ultimately, planets.
The decay products of short-lived radionuclides, those with
half lives comparable to planet forming timescales, time stamp the
evolution of the early solar system.
Several short lived radionuclides appear to be abnormally
abundant and provide critical clues to the environment of the
solar birthplace. I describe these in \S\ref{sec:meteoritics}.

By providing fine detail on small scale processes within one
solar system (ours), cosmochemical studies are complementary to
astronomical observations that show the gross properties
of many protostars and exoplanetary systems.
Perhaps not surprisingly, given the enormous difference in
scales and the intrinsic diversity of star and planet formation,
the two fields do not always agree.
Nevertheless, reconciling the different lines of evidence,
as best as possible, has value in showing the limits of our
current understanding and in placing our solar system into context
with the hundreds of others now known to exist.
In \S\ref{sec:external}, I show how the inferred abundance
of the short lived radionuclide, \fe60,
in small primitive rocks forces us to consider the processes of star formation
in the setting of the largest scales in the Galaxy, that of spiral arms.
This paper is intended for a general audience. At about the same time,
a more specialized astrophysical review of some of the topics raised
here was being independently written and is currently in press
\cite{2010arXiv1001.5444A}.

\section{Star and planet formation}
\label{sec:starform}
\subsection{The interstellar medium}
\label{sec:ism}
Stars form from the gravitational collapse of the interstellar medium.
This consists of gas, about 90\% hydrogen and 10\% helium by number,
with trace quantities of heavier elements which are mostly locked up
in submicron sized dust grains. The dust accounts for about 1\% of the
mass of the interstellar medium and is produced mainly in the winds from
evolved stars and, to a lesser extent, supernovae.

The astonishing precision to which I alluded to previously concerning
the Sun does not apply to the interstellar medium. Whereas we know
each of the Sun's mass, size and age to at least 4 significant figures,
the properties of the interstellar medium are far less precisely
measured with uncertainties of a factor of 2 or even more.
This is due to the nature of the interstellar medium
as a diverse collection of diffuse, inhomogeneous clouds with
irregular (possibly fractal) boundaries and uncertain distances
that are subject to varied physical influences and are generally
far from thermodynamic equilibrium \cite{2005pcim.book.....T}.
Nevertheless, the range of scales involved in star and planet formation
are so vast that these rough measures are sufficient to elucidate the
dominant physical mechanisms at work.

The interstellar medium exists in a wide range of temperatures, $T$,
and number densities, $n$, in distinct phases in approximate pressure
balance, $P=nkT\approx 10^{-13}$\,Pa \cite{2005ARA&A..43..337C}.
Most of the volume consists of hot, rarefied, ionised gas with
$T\approx 10^6\,{\rm K}, n\approx 5\times 10^3$\,m\eee,
heated by supernovae shock waves.
Most of the mass is in atomic clouds with a range of temperatures
$T\approx 10^2-10^4$\,K and densities $n\approx 50-0.5\times 10^6$\,m\eee\
respectively.
The coldest, densest regions of the interstellar medium are molecular
with $T\approx 20\,{\rm K}, n\simgt 5\times 10^8$\,m\eee,
Because of their cool temperatures, the peak of the emission occurs
at a wavelength $\lambda=hc/kT\simeq 1$\,mm, and high frequency radio
observations are required to observe the emission from this phase.

Molecular clouds come in a range of sizes and are the sites
of all star formation. Small clouds dominate by number
but large clouds by integrated mass. Using astronomical units
for a solar mass, $1\,\Msun=2\times 10^{30}$\,kg, and parsec,
$1\,{\rm pc}=3.09\times 10^{16}$\,m, {\it giant molecular clouds}
are those with masses
$M\approx 10^4-10^6$\,\Msun\ and sizes $L\approx 10-100$\,pc.
These are the most massive objects in the Galactic disk and
are where most stars are born.

Each phase of the interstellar medium is inhomogeneous but molecular
clouds are the most structured with a hierarchy that is variously
described as {\it clumps} and {\it cores} \cite{2000prpl.conf...97W}.
Clumps range from sizes $L\approx 0.5-5$\,pc
and masses $M\approx 10-10^3$\,\Msun\
and may contain young groups, or {\it clusters}, of stars.
Cores are the smallest identifiable structures that form individual
stars and are approximately spherical with typical radii
$R\approx 0.05$\,pc and mass $M\approx 1$\,\Msun\
\cite{2007prpl.conf...17D}.

\begin{figure}[ht]
\includegraphics[width=6.3in,angle=0]{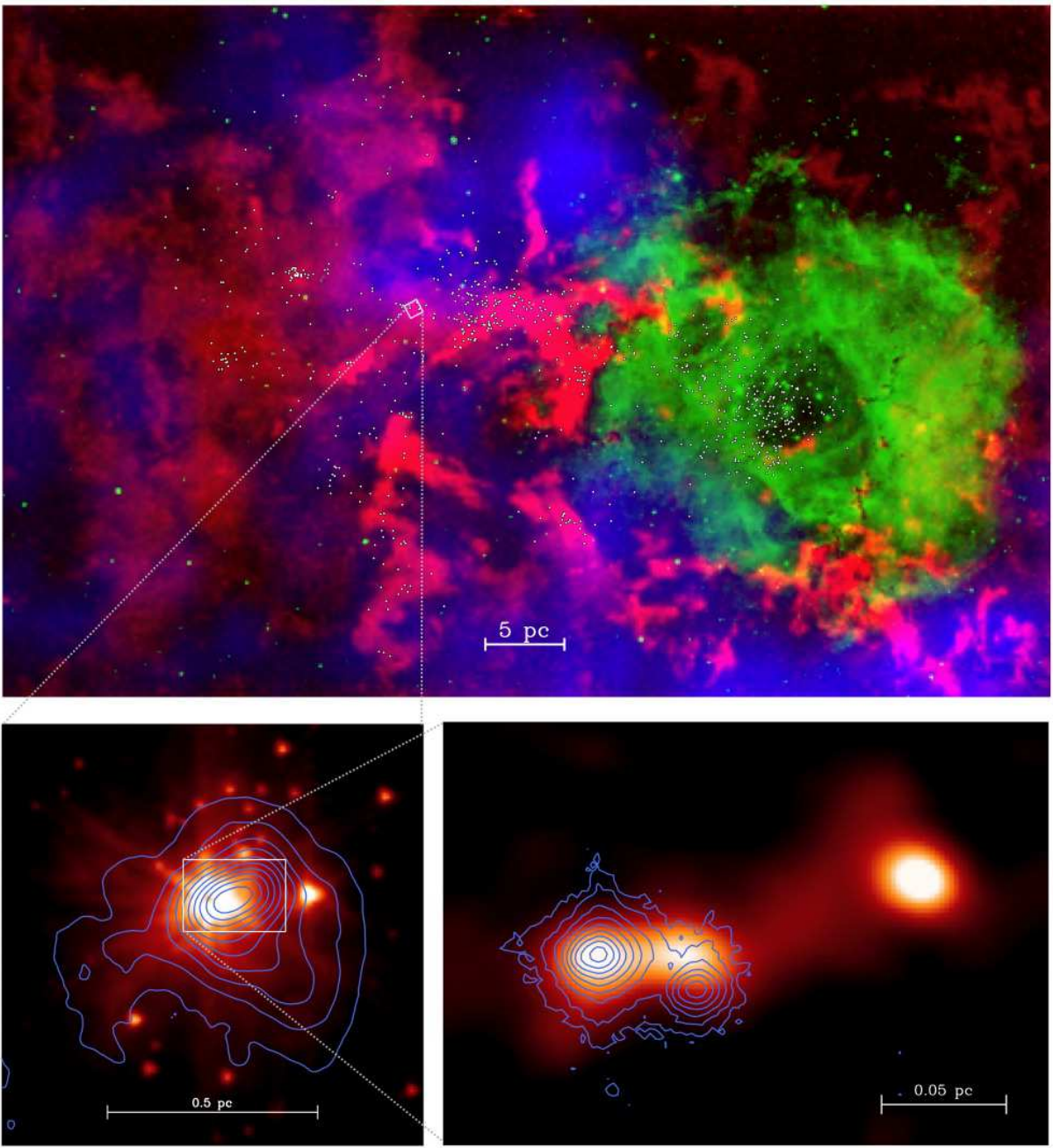}
\caption{The scales of star formation. The upper panel shows a composite
view of the Rosette nebula and accompanying cloud. The nebula, shown in
green from a Digitized Sky Survey image, is powered by a collection
of massive stars at the centre of a large cluster.
The red image is 2.6\,mm emission from
the CO molecule and indicates the presence of a Giant Molecular Cloud,
about 50\,pc in diameter and mass $2\times 10^5$\,\Msun\
\cite{2006ApJ...643..956H}.
The blue image shows the intensity of the neutral hydrogen 21\,cm line,
and reveals an atomic envelope around the cloud \cite{1993ApJ...414..664K},
either the remnants of the diffuse gas from which the molecular
cloud formed or photodissociated gas due to the surrounding ultraviolet
radiation field.
The lower panels zoom in on the star formation process. The left panel
shows contours of 850\,\micron\ emission from a cold dusty envelope
around a deeply embedded cluster, imaged at 3.6\,\micron.
The right panel shows 4.9\,\micron\ contours from
embedded protostars on a 1.2\,mm interferometric image of
3 dusty cores lying in a filament at the centre of the clump
\cite{2009ApJ...699.1300W}.
The scale bars decrease by an order of magnitude in each panel
demonstrating the enormous change in scale from cloud to clump to core.}
\label{fig:clouds_clumps_cores}
\end{figure}

Figure~\ref{fig:clouds_clumps_cores} illustrates the hierarchy
and huge range of spatial scales involved in the full star formation
process from diffuse atomic gas to giant molecular cloud to
cluster forming clump to individual star forming core.

The velocity dispersion of molecular clouds, $\sigma\approx 3-10$\,\kms,
is much larger than the sound speed,
$\sigma_{\rm thermal}=(kT/\mH2)^{1/2}=0.1$\,\kms\
where $\mH2$ is the mass of a hydrogen molecule,
and is likely due to waves along magnetic field lines.
Consequently, unlike the ionised and atomic components of the
interstellar medium, molecular clouds are not in pressure equilibrium
with their surroundings.
Either they are bound by their own self-gravity or they are transient
features rather like the froth on an ocean wave.
This is a case where the lack of precise measurements hinders our
ability to differentiate between these two possibilities and it is
hard to determine the lifetime of molecular clouds
(or even tell the difference between young and old objects
\cite{1996ApJ...464..247W}).
Current estimates, based on theories of cloud formation and destruction,
range from $\approx 3-10$\,Myr
\cite{2001ApJ...562..852H,2002ApJ...566..302M}.
The evolutionary state of the denser subunits, clumps and cores,
within star forming clouds are easier to characterize and several
lines of evidence suggest that stars form within a few gravitational
free-fall timescales,
\begin{equation}
\label{eq:freefall}
t_{\rm ff}=\left(\frac{3\pi}{32G\rho}\right)^{1/2},
\end{equation}
where $\rho$ is the mass density \cite{2009ApJS..181..321E}.
For a number density $\nH2\approx 10^{10}$\,m\eee, $t_{ff}\approx 10^5$\,yr

\subsection{Protostars and protoclusters}
\label{sec:protostars}
As the magnetic or turbulent support of a dense molecular core
is lost, gravitational collapse ensues. The release of gravitational
energy produces a hot central object that is observable at
infrared wavelengths. The onset of hydrogen fusion at a core
temperature of $6\times 10^6$\,K, signals the birth of
of a {\it protostar}.
The radiation and winds from the young star sweep away
any remaining molecular material and it becomes optically
visible \cite{2005fost.book.....S}.

Ground-based astronomy can detect protostars at
$\lambda\simlt 20$\,\micron\ but space-based instruments are required to
observe cooler and fainter objects at longer wavelengths.
The Spitzer and Herschel telescopes are now providing detailed
views of the youngest protostars in molecular clouds.
Infrared surveys of molecular clouds show that most stars
do not form singly but in clusters ranging from
10 to $10^6$ members \cite{2003ARA&A..41...57L}.

The number of clusters, $N_{\rm c}$ with $N_*$ stars follows a
differential equation of the form
\cite{2003ARA&A..41...57L,2001AJ....122.3017S}.
\begin{equation}
\label{eq:dNc}
\frac{dN_{\rm c}}{dN_*} \propto N_*^{-2},
\end{equation}
which implies that the probability that any given star forms in
a cluster with $N_*$ stars is
$P(N_*)=\int N_* dN_{\rm c} \propto \log\,N_*$.
That is, a star is as likely to be born in a rare, but large
cluster with between $10^5$ and $10^6$ stars as 
in a far more common smaller group with between 10 and 100 stars.
Even though our Sun is an isolated system today, it most likely 
formed with many, likely thousands of, siblings that have since
dispersed during their twenty-plus orbits around the Galaxy.

Stars are born with a range of masses, $M\approx 0.1-100$\,\Msun,
with a distribution that appears to be
remarkably uniform \cite{2001MNRAS.322..231K}.
The number of stars with mass $M_* \simgt 0.5$\,\Msun\ follows
a power-law distribution,
\begin{equation}
\label{eq:IMF}
\frac{dN_*}{dM_*} \propto M_*^{-2.35}.
\end{equation}
Only a small fraction, $\approx 0.3$\%, are massive
enough, $M_*>8$\,\Msun, to become core-collapse supernova.
The distribution is heavily skewed toward the low end in terms
of numbers but the few massive stars dominate the luminosity and their
radiation, winds, and eventual supernovae affect the
interstellar medium thermally, dynamically, and chemically.

\subsection{Protoplanetary disks}
\label{sec:disks}
The huge compression from core to stellar scales
($\approx 10^{15}\rightarrow 10^9$\,m) magnifies any initial
spin in the cloud and results in a rotationally supported {\it disk}
around a protostar. These disks initially funnel material onto the growing
star but they are also the sites of planet formation, being dense enough
for dust grains to aggregate and, in their midplane, cool enough for
gaseous compounds such as water to freeze out.

Disks are warm near their central star and cool far away,
$T\approx 10^3-10$\,K.  Consequently they
radiate from infrared to millimeter wavelengths.
Infrared surveys are the most sensitive indicators of the presence of
a disk and large surveys of clusters with different ages
show that the fraction of stars with disks decreases from an initial
value close to unity to zero by by 6\,Myr, and that the median disk lifetime
is about 3\,Myr \cite{2001ApJ...553L.153H,2008ApJ...686.1195H}.
Masses can be measured at millimeter wavelengths where the entire disk emits.
Observations show disk masses ranging from $10^{-4}$ to $10^{-1}$\,\Msun,
around stars with ages $\simlt 1$\,Myr.
This brackets the mass required to form the planets
in the Solar System, estimated to be 0.01\,\Msun\
\cite{2005ApJ...631.1134A,2007ApJ...671.1800A}.
However, the millimeter emission decreases precipitously
by 3\,Myr \cite{2002AJ....124.1593C}.
indicating that the emitting surface area has decreased,
either due to disk dispersal or grain growth into relatively
large (centimeter or greater) sized {\it planetesimals}.

Disks are small with initial radii about 100 times the Earth-Sun
distance, $1\,{\rm AU}=1.5\times 10^{11}\,{\rm m}\approx 5\times 10^{-6}$\,pc.
They can only be imaged at millimeter wavelengths using the technique
of interferometry, that of correlating the signals from multiple antennae.
In the case of Orion, however, disks are seen silhouetted against the bright
backdrop of the nebula in optical light. The Hubble Space Telescope
images of these so-called proplyds are some of the most spectacular images
of protoplanetary disks \cite{1998AJ....116..293B,1996AJ....111.1977M}.
About 15\% of disks in young star forming regions have the
combination of sufficient mass $M\geq 0.01$\,\Msun\ within $R\leq 50$\,AU
to form a planetary system on the same scale as our own
\cite{2009ApJ...694L..36M}.

\section{Meteoritics}
\label{sec:meteoritics}
\subsection{Chondrites}
\label{sec:chondrites}
Meteorites provide a unique way to infer the conditions of the early
solar system in far greater detail than astronomical measurements of
a distant protoplanetary disk. These rocks are the building blocks
of planets that, until they landed on Earth,
were not incorporated into a large solar system body.
As such they are fossils of the early stages of planet formation.
Using scanning electron microscopes and ion mass spectrometry
cosmochemists can measure the structure, mineralogy and isotopic
composition of meteorites to learn about the precise conditions of
the environment in which they formed.

Meteorites can be stony, iron, or a mixture of the two.
The most primitive are stony with a composition similar to that
of the Sun. These so-called {\it chondrites} are themselves largely
composed of {\it chondrules}, approximately spherical globules generally
less than a millimeter across,
made largely of silicates and small amounts of metals.
Carbonaceous chondrites are a subclass with large abundances of
refractory elements, those that condensed out of the protosolar nebula
at relatively high temperatures, and therefore early in its history.
In particular, carbonaceous chondrites contain many Calcium-Aluminium-rich
inclusions,  the oldest known solids in the early solar system
with an age derived from radioactive lead isotopes to
be $4567.2 \pm 0.6$\,Myr \cite{2007prpl.conf..835W}.

Radioactive isotopes with much shorter half-lives show the
time sequence of important events in the growth of planets
and can be directly compared to astronomical observations
of Myr-age protoplanetary disks.

\subsection{Short lived radionuclides}
\label{sec:slr}
An element can have unstable isotopes that are chemically identical but
decay over cosmic timescales. A chondrule with a particular mineralogy
can therefore be incorporated into a chondrite but subsequently change
its composition. These isotopic anomalies are frozen into the material
and, unless the chondrite undergoes further processing, are immutable.
Short lived radionuclides have half-lives, $t_{1/2}$, comparable to planet forming timescales,
variously defined as $10-100$\,Myr \cite{2005ASPC..341.....K}.
and provide a natural chronometer for the growth of planets.

The first short lived radionuclide, $^{129}$I, with a mean life
($\tau=t_{1/2}/\ln\,2$)
of 23.5\,Myr was discovered in meteorites almost 50 years
ago \cite{1961JGR....66.3582J}.
Many others have since been detected \cite{ErnstZinner04112003}.
Their importance lies in showing the timescales over which the
products of stellar nucleosynthesis are transported through the
interstellar medium to planetesimals.
The total number of atoms of a radionuclide depends on a balance
between production rate, $P$, from stellar nucleosynthesis
and radioactive decay \cite{1970ApJ...162...57S},
\begin{equation}
\label{eq:slr1}
\frac{dN}{dt}=P-\frac{N}{\tau}.
\end{equation}
For a stable isotope, $\tau$ is effectively infinity, and $N=\int Pdt$,
the production rate integrated over the age of the Galaxy.
For a finite mean life, the general solution to the above is
\begin{equation}
\label{eq:slr2}
N=\int_0^tP(t')e^{-(t-t')/\tau} dt'.
\end{equation}
The exponential within the integral strongly weights the
number toward the recent production history (i.e., within a few mean lives).
The predicted ratio of an short lived radionuclides relative to
its stable partner in a chondrule must take into account the additional
decay during the transport from hot, rarefied stellar ejecta to solid.
Measured values are roughly consistent with a timescale of about
100\,Myr but there is a significant discrepancy from one element
to another that may be attributable to different production rate
histories \cite{2005ASPC..341..548J,2009GeCoA..73.4922H}.

\begin{figure}[ht]
\includegraphics[width=5.5in,angle=0]{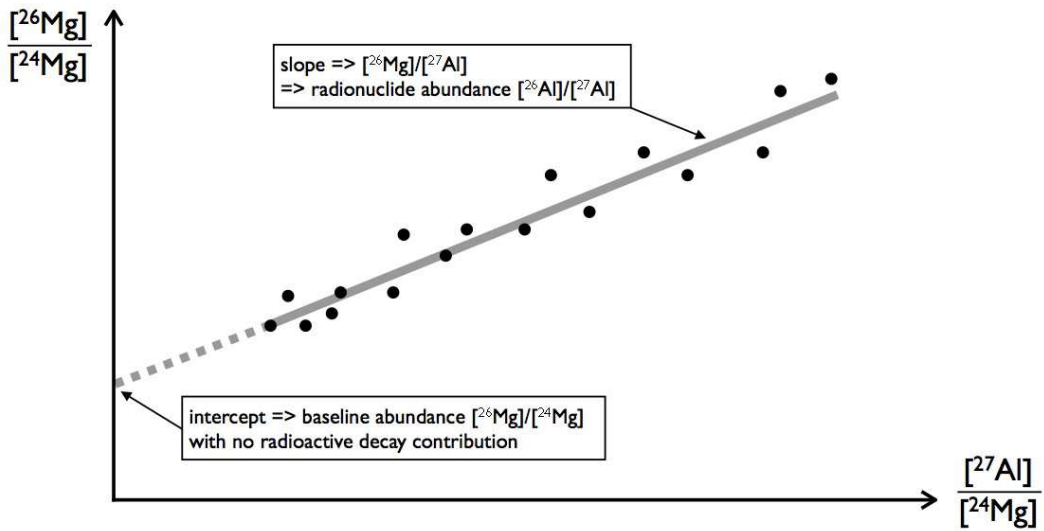}
\caption{Schematic of the determination of the \al26\ isotopic
abundance in Calcium-Aluminium inclusions.}
\label{fig:al26}
\end{figure}

A few very short-lived short lived radionuclides with
$\tau < 3$\,Myr stand out as having abnormally high abundances.
An important example is \al26, which decays into $^{26}$Mg with a
mean life $\tau=1$\,Myr. $^{26}$Mg is stable and found in
Calcium-Aluminium inclusions with an abundance that correlates
with $^{27}$Al. The implication is that some of the $^{26}$Mg results
from the decay of \al26.
The situation is shown schematically in Figure~\ref{fig:al26}.
The extrapolation to zero aluminium content shows the baseline
value of $^{26}$Mg and the slope shows the contribution due
to \al26\ decay.
The inferred isotopic ratio in the early solar system,
$^{26}$Al/$^{27}$Al=$5\times 10^{-5}$,
is more than an order of magnitude greater than the predicted steady
state Galactic background \cite{1977ApJ...211L.107L}.

The \al26\ Galactic background can be measured directly via
$\gamma$-ray spectroscopy of its decay \cite{2006Natur.439...45D}.
The resolution of these instruments is too low to show
much detail although the Galactic rotation can be
discerned showing that the emission is indeed coming
from throughout the Galaxy.
Compared with these measurements the inferred abundance of \al26\
in the early solar system is a factor of 6 above the Galactic background.
The inescapable conclusion is that there was a significant spike in
the local production rate of \al26\ slightly before or during the
formation of the solar system.

The discoveries of short-lived radionuclides in chrondrites not only
provide important constraints to the solar birthplace environment
but also have wider ramifications.
The rapid decay and high abundance of \al26\ produces enough heat to melt
and differentiate planetesimals at early times \cite{2007M&PS...42.1529S}.
Such radiogenic heating may significantly move the {\it snow line},
the disk radius at which water freezes, outwards in a forming planetary
system \cite{2004LPI....35.1987D}.
Thus the issue of the commonality of the environment of the early
solar system, is of great import for understanding the water content
of exoplanets and even their capacity to host technological
civilizations \cite{2009Icar..201..821G}.
Too little water would be inhospitable to life but too much water,
covering all the land, would prohibit technological development.
Although astronomers are only just at the cusp of being
able to determine the composition of exoplanets, one ocean-world
has now been convincingly demonstrated from radial velocity
measurements of its mass and transit measurements of its radius
\cite{2009Natur.462..891C}.
Such density determinations provide a direct link between
astronomical and cosmochemical studies.

\subsection{An external influence on the early solar system}
\label{sec:60fe}
Two explanations were proposed to explain the high abundance of \al26.
An external enrichment of the protosolar nebula by the winds or
supernovae of nearby massive stars \cite{1977Icar...30..447C}
or a local production by the irradiation of chondrites by energetic
particles from an active protosun \cite{1998ApJ...506..898L}.

External enrichment requires a synchronicity both in time, due to
radioactive decay, and space, due to geometric dilution of the stellar ejecta.
This led to the idea that, in some way, the death of a massive star
somehow induced the birth of the solar system through, perhaps, the
compression and subsequent collapse of a molecular core.

As our understanding of star formation increased, it was recognised
that young protostars possess very strong magnetic fields and
exhibit strong X-ray flares. The flares are due to magnetic
reconnection events on the stelar surface and shoot particles
(mostly hydrogen and helium nuclei) away at high energies.
The same phenomenon occurs in solar flares today but at a
much reduced frequency and at lower energies due to the much
weaker present-day magnetic field.  As dust grains accreted
through the protosolar disk, they would have been exposed to a high
particle flux and sufficient nuclear reactions would occur
to explain many of the enhanced short lived radionuclides abundances.
This picture is appealing
in many ways, not the least of which is that it is self-contained
and does not require any special circumstances for the early solar system.

A dichotomy has arisen in the literature as both external and internal
mechanisms have considerable flexibility and can be adjusted to match
the observed isotope abundances in chondrules. They need not be not
mutually exclusive, however, and it appears that both mechanisms are
needed to explain the observations.
Stellar nucleosynthesis is unable to produce $^{10}$Be at the levels
observed in chondrules \cite{2000Sci...289.1334M}.
Its presence is definitive proof that the protosun did mutate
some of the elements in the disk.
On the other hand, protostellar particle fluxes are several orders of
magnitude too low to match the observed abundance of the
neutron-rich short lived radionuclides, \fe60\ \cite{1998ApJ...506..898L}.

\fe60\ decays to $^{60}$Ni with a mean life of 3.8\,Myr
\cite{2009PhRvL.103g2502R}
and its presence in chrondrites was inferred by the
correlation of the isotope ratio, $^{60}$Ni/$^{61}$Ni,
with the main isotope of iron, $^{56}$Fe/$^{61}$Ni \cite{2003ApJ...588L..41T}.
Although the evidence for \fe60\ in chondrites does not appear
to be in dispute, its precise abundance in the early solar system
is not yet firmly established with values ranging from
[\fe60]/[$^{56}$Fe]$ < 10^{-7}$ \cite{Regelous2008330}
to $10^{-6}$ \cite{2005ApJ...625..271M}.
Here, I use an intermediate value,
[\fe60]/[$^{56}$Fe]$=3\times 10^{-7}$ \cite{2006ApJ...639L..87T}
which places it about an order of magnitude greater
than the model background level and a factor of 6 above
direct $\gamma$-ray measurements \cite{2005A&A...433L..49H}.

The elevated abundance of \fe60\ is a clear and unambiguous
signature of external enrichment of the early solar system.
Understanding its transport from the core of a massive star
to planetesimal in at most a few mean lives provides the
strongest constraint on the astrophysical environment of the
solar birthplace. In the following section I describe various
short lived radionuclides enrichment scenarios.

\section{External enrichment scenarios}
\label{sec:external}
Iron and its short lived isotope \fe60\ can only be produced in the cores of
massive stars. It can then be expelled into the interstellar medium either
through stellar winds during the very late stages of their
evolution as core material is convected up to the surface or during
the death of a star as a core-collapse (Type 1b, 1c, or II) supernova.
The fundamental issue is how to convert hot and rarefied stellar winds
or supernova ejecta into dust grains in a cold, dense protoplanetary
disk within a sufficiently short time that the radionuclide does not
decay away. Both sources, stellar winds and supernovae, have historically
been considered in the context of the \al26\ excess. I discuss each below
but now with the requirement to match the cosmochemically inferred
early solar system \fe60\ abundance.

\subsection{Winds from evolved stars}
\label{sec:agb}
\begin{figure}[ht]
\includegraphics[height=6.3in,angle=90]{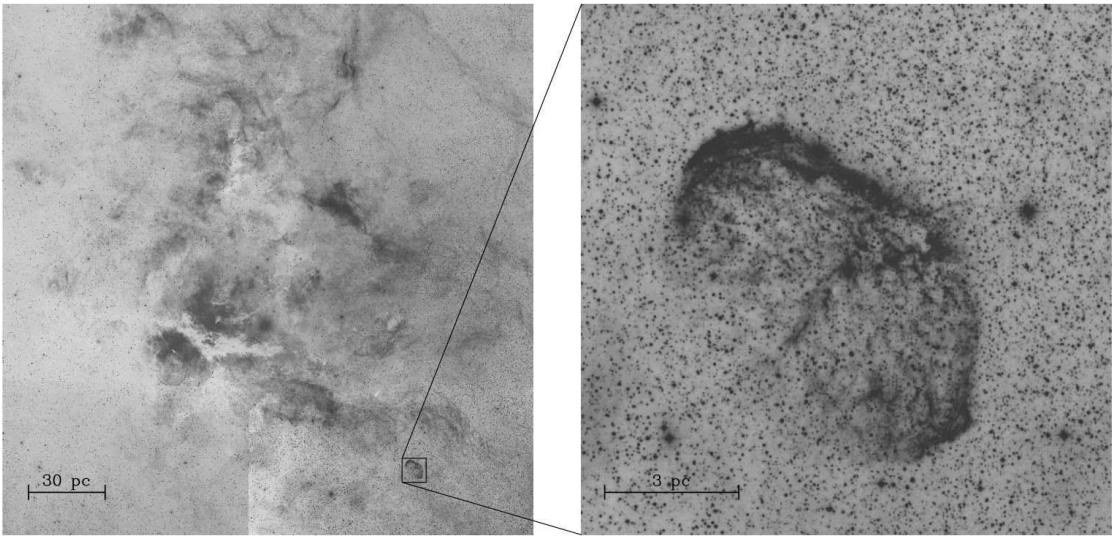}
\caption{The relative scale of stellar wind ejecta compared to molecular
clouds. The two panels are from the Digitized Sky Survey and show optical
extinction and nebulosity associated with the IC1318 star forming
region in the left panel. The right panel shows a close-up of the
wind-blown bubble from the evolved star, HD\,192163. Note the relative
scales that graphically demonstrates how the pollution from a stellar wind
is limited to only a small part of a molecular cloud.}
\label{fig:agb}
\end{figure}

The end state of a star begins when all the hydrogen in its centre is
used up. The core then contracts under its own gravity, heats up,
and shells of hydrogen and helium undergo fusion around it.
The outer layers of the star expand due to the increased energy
production and it becomes a red giant. The large size implies low
surface gravity and substantial gas can escape in the form of a wind.
Convective mixing from core to surface then provides a path for newly
synthesized material to rapidly enter the interstellar medium.
This mechanism is a potential explanation of the high abundances of
many of the short lived radionuclides inferred in the early
solar system \cite{1993prpl.conf...47C}.
Figure~\ref{fig:agb} shows a wind-blown bubble from an evolved
star in a star-forming molecular cloud; a simultaneous demonstration
that this situation can occur but that its effect is limited to a
very small region of the cloud.

The production of \fe60\ requires high core densities that are
only realised in moderately massive stars with masses
$M_\ast\simgt 5$\,\Msun\cite{2006NuPhA.777....5W}.
It takes about 110\,Myr for a 5\,\Msun\ star to enter the
red giant phase \cite{1992A&AS...96..269S},
much longer than molecular cloud lifetimes (\S\ref{sec:ism}).
More massive stars have shorter lifetimes but only the most massive,
$M_\ast\simgt 20$\,\Msun, use up their core hydrogen within the 10\,Myr upper
bound to giant molecular cloud lifetimes and they become supernovae
shortly thereafter. The \fe60\ yield in supernovae dominates in
this case, as discussed in subsequent subsections.

The external enrichment of a planetary system by a red giant therefore
requires the chance coincidence of a relatively old evolved star
running into a molecular cloud just as it begins to form stars.
Based on a census of known red giants and molecular clouds within
1000\,pc of the sun the probability of such serendipity is estimated
to be $\approx 1$\% per Myr \cite{1994ApJ...421..605K}.
Further, the ratio of stellar wind to molecular mass is very small,
$3\times 10^{-4}$, so the likelihood that the first few Myr of the
nascent solar system were enriched with \al26\ at the cosmochemically
inferred levels is about 0.001\%.
The same reasoning applies for the enrichment of \fe60\ but with
an even lower likelihood due to the required high stellar masses.

Catalogs of evolved stars with high mass loss rates extend beyond the
solar neighborhood \cite{1990ApJ...364..663J}
and can be extrapolated to make a more general calculation for the
entire Galaxy. Their average surface density is about $10^{-5}$\,pc\ee,
independent of Galactic radius. This implies a total population
of about $3\times 10^3$ in the Galaxy. Each object remains in this
phase for about 1\,Myr and ejects a total of about 3\,\Msun\ into
the interstellar medium.
Modeling the isotopic abundances in the winds, each stars can pollute
about 100 times more mass to early solar system levels
or $\approx 10^6$\,\Msun\ in total. The total molecular mass in the Galaxy is
$1\times 10^9$\,\Msun\ \cite{1997ApJ...476..166W}
so the proportion of star forming material that can be enriched
is no more than 0.1\%.
This assumes that the star formation efficiency is independent
of the presence of an evolved star which is entirely consistent
with observations.
Note that this is a generous upper limit since it does not
include the low likelihood that an evolved star lie within or
near a molecular cloud, nor does it
include radioactive decay during the transport from stellar wind
to planetesimal. Finally, again no account has been made to 
consider the subset of evolved stars the are massive enough to
produce \fe60.  Even without these additional factors,
however, this simple calculation provides a robust and independent
demonstration that the pollution of the early solar system
by the winds from an evolved star is an extremely unlikely scenario
for the origin of \fe60\ in the early solar system.

\subsection{Supernovae}
\label{sec:supernova}
The most massive stars, $M > 8\,\Msun$, have such high central pressures
that shell burning around a hydrogen depleted core progresses to
more neutron rich elements up to iron, at which point nuclear fusion
is no longer an exothermic process. Without a source of thermal
pressure to counteract self-gravity, the star collapses in on itself.
The release of gravitational potential energy produces a
supernova explosion and many of the freshly synthesized elements
are ejected and can be recycled into a new generation of stars and
planets. Indeed, almost all the iron on Earth was produced in supernovae
but the presence of its isotope, \fe60\ with mean life $\tau=3.8$\,Myr,
ties the death of a massive star to the nearly simultaneous formation
of the solar system.

Compared to stellar winds, supernovae are an instantaneous event but
the huge energy release impacts a much larger volume. The initial
speed of supernovae ejecta is $\approx 10^4$ \,\kms, and
supernovae remnants can expand to hundreds of parsecs in diameter
\cite{1988ARA&A..26..145T}.
Their influence on the different scales of star and planet formation,
however, depends on a number of factors which is incorporated into
the concept of radioactivity distance.

\subsubsection{The radioactivity distance}
\label{sec:raddist}
In the transport of an short lived radionuclide from supernova
to planetesimal, there will be losses due to radioactive decay,
geometric dilution, and efficiency of incorporation into solids.
Assuming that a mass, $M_{\rm SLR}$, is ejected isotropically,
the resulting mass fraction in a target at distance $D$
with mass $M_{\rm t}$ amd projected area $A_{\rm t}$ is,
\begin{equation}
\label{eq:raddist}
X_{\rm SLR}=f_{\rm inj}\frac{M_{\rm SLR}}{M_{\rm t}}\frac{A_{\rm t}}{4\pi D^2}\,e^{-t/\tau},
\end{equation}
where $f_{\rm inj}$ is the injection efficiency \cite{2006ApJ...652.1755L}.

Relative to hydrogen, the solar iron abundance is
[$^{56}$Fe]/[H]$=3.45\times 10^{-5}$ so the
chondritic isotopic ratio [\fe60]/[$^{56}$Fe]$=3\times 10^{-7}$ 
corresponds to a mass fraction $X_{60{\rm Fe}}=4.4\times 10^{-10}$
(the product of the two abundances times the ratio of the mass of
\fe60\ to the mean atomic mass $1.4m_{\rm proton}$ from an admixture
of 90\% hydrogen and 10\% helium).
This is the number that must be matched or exceeded by any
transport scenario.

Stellar nucleosynthesis models yield
$M_{60{\rm Fe}}=2\times 10^{-6}-1\times 10^{-3}$\,\Msun\
for supernova progenitor masses
$M_*=11-120$\,\Msun\ \cite{2006ApJ...647..483L}\footnotemark
\footnotetext{
The mass fraction of \fe60\ relative to $^{56}$Fe is about $10^{-5}$
in supernova ejecta, about two orders of magnitude greater than the
ratio in primitive rocks so the overall iron abundance in the solar
system is increased by only about 1\% from the \fe60\ injection event,
and would not be discernable in the galactic metallicity distribution.
}.
Taking the stellar mass function weighted mean,
$M_{60{\rm Fe}}=1.1\times 10^{-5}$\,\Msun, and assuming a maximum
$f_{\rm inj}=1$ and negligible transport time, $t\ll\tau$,
gives an upper limit to the distance between a source and target
for enrichment of \fe60\ to early solar system levels,
\begin{equation}
\label{eq:d60}
D_{60}{\rm (pc)}=45\,\Sigma^{-1/2}_{\rm t}{\rm (M_\odot\,{\rm pc}^{-2})},
\end{equation}
where $\Sigma_{\rm t}=M_{\rm t}/A_{\rm t}$ is the projected
surface density of the target. This radioactivity distance
serves as a rough guide to the relevant scales of interest.
Four different enrichment scenarios are ordered below by target scale
from smallest (disks) to largest (giant molecular clouds)
ending with what I believe to be the most probable solution,
that of an enhanced background in spiral arms.

\begin{figure}[ht]
\includegraphics[width=6.3in,angle=0]{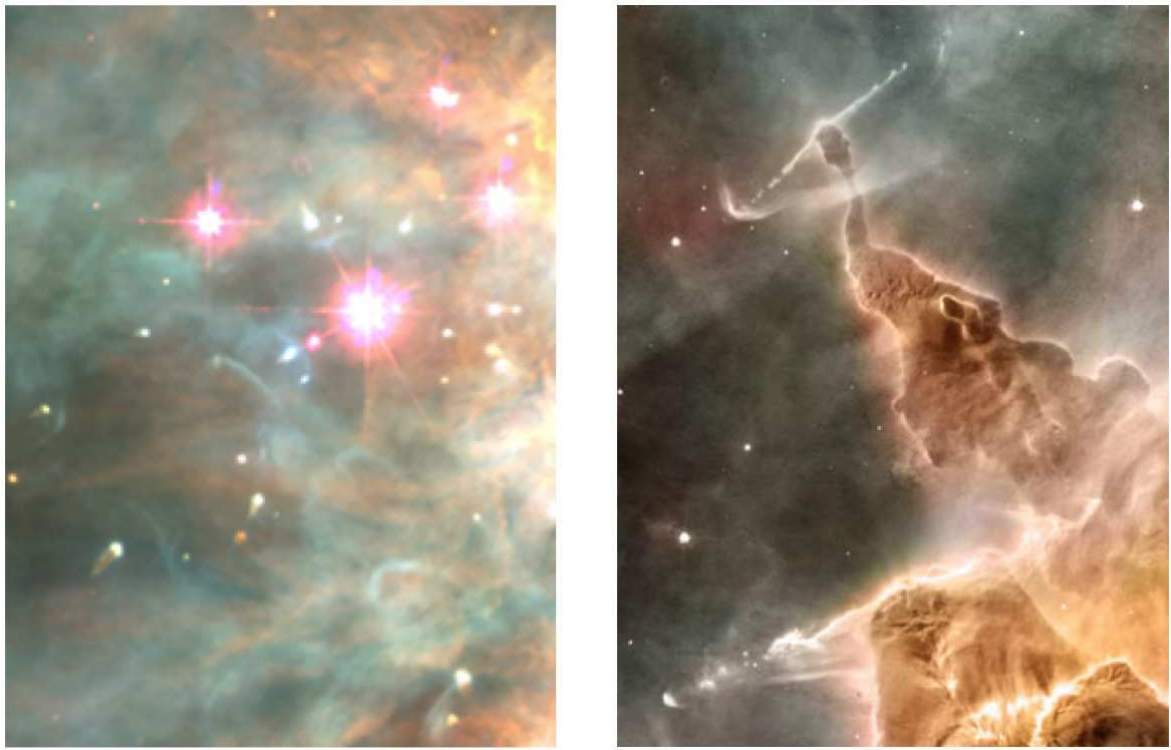}
\caption{Hubble Space Telescope images of star and planet formation
within or on the boundaries of ionized regions around massive stars.
The bright stars in the left panel is the Trapezium Cluster in Orion.
Cometary tails of photoevaporating protoplanetary disks around low mass stars
can be seen pointing back toward the brightest star, $\theta_1$\,Ori\,C.
The right panel shows two star-forming pillars in the Carina nebula.
Jets from young protostars at the tips of the pillars indicate
very recent, potentially triggered, star formation.}
\label{fig:triggered_sf}
\end{figure}

\subsubsection{Injection into a protoplanetary disk}
\label{sec:sndisk}
Clusters extend in size up to about $N_*\approx 10^6$ stars so,
with the cluster and number distributions as described
in \S\ref{sec:protostars}, the median cluster size is $N_*\approx 10^3$
which is sufficient to contain a star massive enough to become a supernova.
Direct evidence that the solar system formed in a large cluster comes from
the sharp edge to the radial distribution of Kuiper belt objects
(rocky bodies beyond the orbit of Neptune) which has been
attributed to the external photoevaporation of the protoplanetary
disk by a nearby hot, ionizing star \cite{1998AJ....115.2125J}.
The Trapezium Cluster in Orion is the closest example of such
a star forming environment (Figure~\ref{fig:triggered_sf}).
As the massive stars end their lives as supernovae, nearby
protoplanetary disks could, in principle, be directly injected
with \fe60\ \cite{2007ApJ...662.1268O}.

For a disk with the minimum mass necessary to form
the solar system, $M=0.01$\,\Msun, in a radius $R=50$\,AU, 
the face-on surface density, $\Sigma=5\times 10^4$\,\Msun\,pc\ee\
implies a radioactivity distance $D_{60}=0.2$\,pc.
This is on the limit at which a protoplanetary disk can
survive a supernova explosion \cite{2000ApJ...538L.151C}
which immediately suggests that it is an unlikely scenario.

\begin{figure}[h]
\includegraphics[height=6.4in,angle=90]{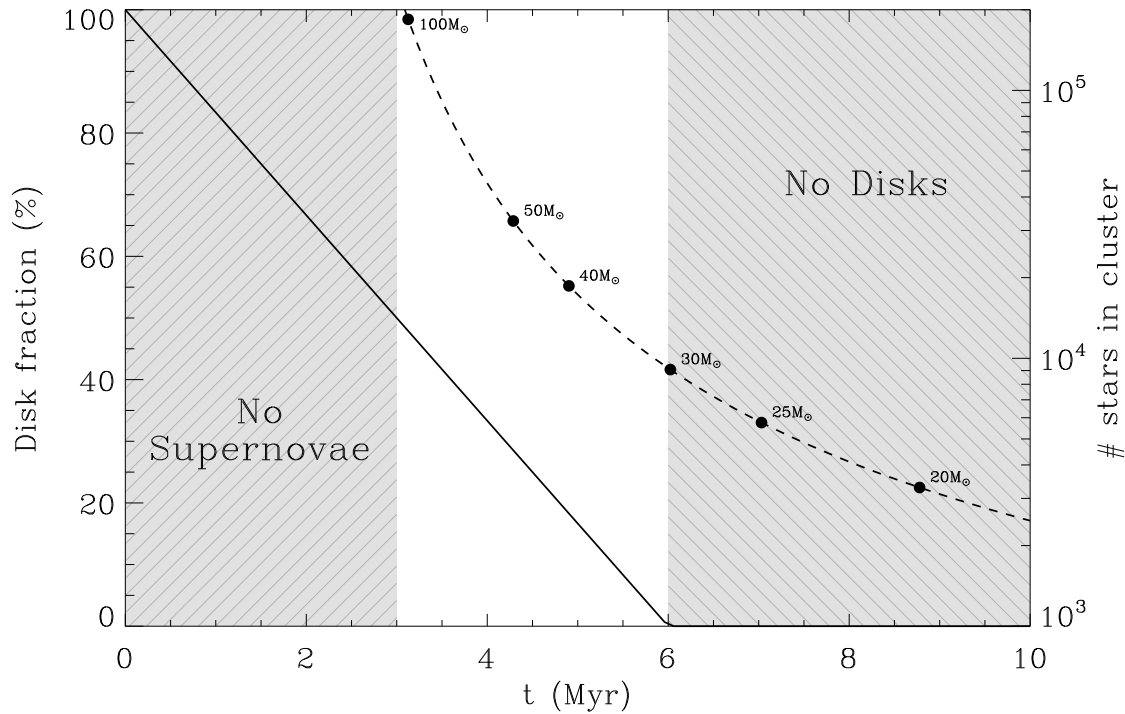}
\caption{The lifetime problem for direct supernova injection
into a protoplanetary disk.
The solid diagonal line decreasing from 100\% to 0\%
schematically represents the decrease in the disk fraction
in clusters of varying age \cite{2001ApJ...553L.153H}.
The curved dashed line plots the expected cluster size
required to host a massive star with main sequence lifetimes
from 3 to 10\,Myr. Stars of different mass are labeled along
this line from $20-100$\,\Msun.
The hashed areas show that there are no supernovae before 3\,Myr
and no disks remain after 6\,Myr. This leaves only a small range
in time, $3-6$\,Myr, when there are both disks and a supernova.
Moreover, only stars with masses greater than
30\,\Msun\ satisfy this constraint and they are extremely rare,
found only in clusters of $10^4$ stars or more.}
\label{fig:life}
\end{figure}

An additional problem is the timescale for massive stars to become supernovae
compared to disk lifetimes. Even the most massive stars,
$M_\ast = 100$\,\Msun, take about 3\,Myr to burn their core hydrogen
and begin nucleosynthesis of heavy elements at which point about
half the disks in the cluster will have disappeared.
Further, only stars with masses $M_\ast > 30$\,\Msun\ explode within 6\,Myr,
the maximum disk lifetime. Stars this massive are extremely rare and only
found in very large clusters, $N_*\simgt 10^4$.
The situation is graphically illustrated in Figure~\ref{fig:life}.

The timescale problem is only slightly mitigated by allowing the
massive star to form before the low mass stars since the main
sequence lifetime increases sharply as the stellar mass decreases.
For instance, the lifetime of a 20\,\Msun\ star is about 9\,Myr
but this is still a relatively rare star, found only in clusters
with $N_\ast > 3000$, somewhat larger than Orion.
For disks to survive at the time this becomes a supernova would require
that it get a head-start of at least 3\,Myr in its evolution but
there is no evidence that massive stars form that far before other
stars in a cluster and, indeed, the situation appears to be
quite the opposite \cite{2007ARA&A..45..481Z}. 

Modeling cluster evolution and taking into account both the spatial
and temporal constraints shows that no more than 1\% of stars in an
Orion-like cluster with $N_\ast\approx 10^4$ match the \fe60\
solar system abundance \cite{2007ApJ...663L..33W,2008ApJ...680..781G}.
The probability is lower in smaller
clusters since they would be unlikely to host a supernova before
all the disks disappeared and also lower in larger clusters since
most disks would lie beyond the radioactivity distance.
Allowing for multiple supernovae increases the odds and favors the
largest clusters but,
integrating over the full cluster distribution in equation~\ref{eq:dNc},
the probability that any given star in the Galaxy
was enriched with \fe60\ by the direct injection of supernova ejecta
into a disk is a statistically rare event, $<1$\%.

\subsubsection{Triggered star formation}
\label{sec:trigger}
Some stellar associations in molecular clouds consist of spatially
and kinematically distinct subgroups ordered in age.
The canonical example is the four Orion groups, OB1a--d,
spaced sequentially from north to south and with ages
from $\approx 10$ to $\approx 1$\,Myr respectively \cite{1991psfe.conf.....L}.
Such arrangements led to the idea of triggered collapse of molecular gas on
the boundary of an expanding ionised (\HII) region \cite{1977ApJ...214..725E}.
Images of young stars on the brightly illuminated edges of an \HII\ region
abound in the literature in part because of their visual appeal,
e.g., Figure~\ref{fig:triggered_sf}, but the number of protostars is
generally far less than those at the center of the \HII\ region
and it is unclear how representative this mode of star formation
really is.

Nevertheless, if the formation of the Sun was indeed triggered,
the protosolar nebula may have inherited the fresh products of
nucleosynthesis from the massive stellar winds and subsequent supernovae
\cite{2005ASPC..341..107H}.
There is much work on the observations and theory of triggered star
formation \cite{2007IAUS..237.....E} but there are two main mechanisms:
the compression of a molecular cloud core to sufficiently high pressures
that it becomes gravitationally unstable and undergoes collapse,
and the sweeping up of gas to unstable densities and subsequent collapse.

A possible example of the former is the famous ``pillars of creation''
in the Eagle Nebula\cite{1996AJ....111.2349H}.
For a typical core with $M=1$\,\Msun, $R=0.05$\,pc,
$\Sigma=130$\,\Msun\,pc\ee, the radioactivity distance $D_{60}=4$\,pc.
This can be refined by numerical models of supernova ejecta impacting a core
\cite{2002ApJ...575.1144V}
showing that $f_{\rm inj}\approx 0.1$ which implies a reduced $D_{60}=1.2$\,pc.
This is comparable to the distance from the ionizing star
to the pillars in the Eagle nebula which is encouraging.

This picture is deceptive, however, in that the ionizing star
is many Myr away from becoming a supernova.
Well within that timeframe, the powerful stellar winds and
photoionization will sweep away and ablate any cores
\cite{2006ApJ...647..397M, 2003ApJ...594..888F}.
This is consistent with millimeter wavelength observations
that show very little molecular gas toward the centre of ionised or
wind-blown stellar bubbles \cite{2000A&A...357.1001L,2010ApJ...709..791B}.

A final and fatal blow to this scenario has been struck with the most recent
high resolution numerical simulations that show successful triggering
models have very low injection efficiencies,
$f_{\rm inj}\approx 10^{-3}$ \cite{2010ApJ...708.1268B}.
This reduces $D_{60}=0.1$\,pc which is so close to a supernova that
the core would almost certainly be destroyed.
Even if somehow, the ejecta were slowed by dense intervening
gas before impacting the core, the volume argument that showed
a low enrichment likelihood for supernova injection of disks
in \S\ref{sec:sndisk} would apply to cores too.

Triggered star formation may also occur at scales larger
than cores by sweeping up surrounding molecular material
to high densities and subsequent collapse \cite{2009A&A...496..177D}.
The winds and supernova ejecta from the previous generation of stars
may then enrich any resulting planetary systems with some short lived
radionuclides at levels comparable to those in the early solar
system \cite{2009ApJ...696.1854G}.
In this scenario, the relevant scales are those of clumps with typical masses
$M\approx 10^3$\,\Msun\ and (irregularly shaped) sizes $L\approx 5$\,pc.
This implies surface densities, $\Sigma=M/L^2\approx 40$\,\Msun\,pc\ee,
and a radioactivity distance $D_{60}\approx 7$\,pc.

Relative to the pressure driven collapse of a dense core, however,
the ``collect-and-collapse'' of a clump is slow and the
delay between injection of supernova ejecta and star formation
results in significant decay of \fe60.
The low density that makes clumps good targets for injection
also means they have a long free-fall timescale.
The volume density $\rho\approx M/L^3 \approx 5\times 10^{-22}$\,g\,cm\eee,
which implies a free-fall timescale $t_{\rm ff}\approx 3$\,Myr.
The decay term, $e^{-t_{\rm ff}/2\tau}$, therefore becomes significant.
The radioactivity distance reduces to $D_{60}\approx 5$\,pc,
which is smaller than the radii of mature \HII\ regions,
such as the Rosette nebula in Figure~\ref{fig:clouds_clumps_cores}.
The injection efficiency may be higher in lower density clumps
than in cores or disks but it is still likely to be substantially
less than unity, reducing $D_{60}$ still further.

\begin{figure}[ht]
\includegraphics[height=6.3in,angle=90]{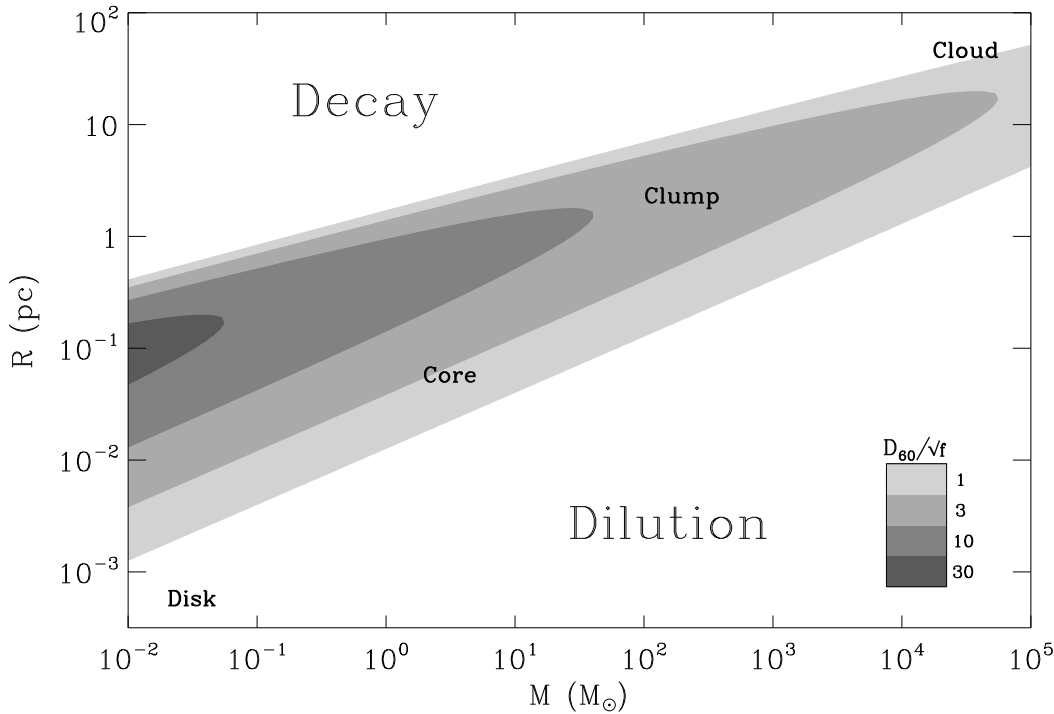}
\caption{Contours of \fe60\ radioactivity distance as a function of
the mass and radius of the enriched object.
Objects with high surface densities cannot capture enough ejecta to
match early solar system values. This region is labeled "Dilution".
Objects with low volume densities have free-fall timescales
much greater than the \fe60\ mean life. This region is labeled "Decay".
The approximate location of the different regimes of star formation,
from clouds to disks, are labeled. The injection efficiency is included
as a scaling to the contour levels, $D_{60}\propto f_{\rm inj}^{1/2}$.}
\label{fig:raddist}
\end{figure}

Figure~\ref{fig:raddist} illustrates the different regimes in a contour
plot of radioactivity distance in $M-R$ space where we have
now allowed for both mass and geometric dilution, $D_{60}\propto\Sigma^{-1/2}$,
and radioactive decay during free-fall, $D_{60}\propto e^{-t_{\rm ff}/2\tau}$.
The injection efficiency is incorporated as a universal scale on the
absolute contour levels, $D_{60}\propto f_{\rm inj}^{1/2}$,
independent of $M$ and $R$.

This figure assumes spherical objects,
i.e., $\Sigma=M/\pi R^2, \rho=3M/4\pi R^3$.
Different geometries produce only minor changes.
Equation~\ref{eq:raddist} also implicitly assumes the supernova ejecta are
isotropically distributed. In fact, observations of supernova remnants
show a great deal of inhomogeneity \cite{2004ApJ...615L.117H}.
In principle, this allows enrichment of a smaller number of objects
(disk, core or clump) at greater distances than $D_{60}$.
The net effect is to decrease the chance of enrichment for any
given object.  Specifically, let the volume filling factor be $f_v$.
Then the ejecta are spread over an area $4\pi D^2f_v$ at distance
$D$ from the supernova. Compared to the isotropic case, any object
along this path can be enriched at a greater distance, $D=D_{60}f_v^{-1/2}$,
but, on average, the number of objects that are affected is smaller
by a factor $f_v$ because they have no preferred distribution
relative to the supernova blast wave.
The overall number of enriched objects is the product of these and therefore
proportional to $f_v^{1/2}<1$.
As a concrete example, consider the extreme case where all the ejecta
are shot out in a narrow beam enriching one object very far from the supernova
but leaving all other objects unaffected.

Figure~\ref{fig:raddist} also shows that the decay problem is even worse at
cloud scales where their low densities imply very slow evolution.
The final mechanism bypasses this issue by effectively creating a
long-lived but localised enriched background within which cloud and
star formation take place.

\subsubsection{Supernova induced cloud formation}
\label{sec:SPACE}
With the growing realization that the conditions for \fe60\
enrichment are unlikely to be due to the direct injection of supernova
ejecta into protoplanetary disks or protostellar cores,
the possibility that supernova ejecta compress their surroundings and induce
the formation of a molecular cloud was proposed \cite{2009ApJ...694L...1G}.
Large clusters contain many massive stars that will, because of their
different masses, produce multiple supernova at a cadence of every few Myr.
This maintains the enrichment of the cloud as it forms a second
generation of stars.
This Supernova Propagation and Cloud Enrichment (SPACE) model
has the advantage that short lived radionuclides are injected
into relatively low density gas over many Myr and should therefore
have a relatively high efficiency of incorporation into star and
planet forming material. It is also appealing in that it can produce
large numbers of \fe60 -rich systems over the entire cloud without
any spatial preference. The conditions of the early solar system would
then be expected to be commonplace.

A critical component of the model is the dynamic conversion of hot
rarefied supernova ejecta into cool atomic and then molecular material via
the collision of large scale shocks and flows in the interstellar medium
\cite{2001ApJ...562..852H}.
If a large stellar association, $N_*=5000$,
remains nearby over this time, about 10 supernova will occur over
20\,Myr and provide an approximately constant flux of \fe60\
into the cloud.

In fact the SPACE model may apply on larger scales and more generally
due to the organization of molecular clouds and massive stars in the
spiral arms of the Galaxy.
Diffuse gas is compressed upon entry into a spiral density wave
allowing, first molecules, and then stars to form.
Almost all short-lived massive stars are born in spiral arms
and many explode as supernova before they can migrate away
\cite{1976ApJ...204..519M}.
This suggests that spiral arms are naturally enriched in short
lived radionuclides, including \fe60.

In classical spiral arm theory, molecular clouds form before the stars
and tend to lie upstream of the spiral density wave \cite{1990ApJ...349L..43R}.
Recent numerical simulations of the feedback from supernova on cloud formation
and destruction and the maintenance of spiral arm structure show that
the collision of supernova bubbles can form new clouds and propagate further
star formation, depending on how much momentum is injected within the disk
\cite{2008ApJ...684..978S}.
These would appear to be precisely the conditions required for
the SPACE model but now on spiral arm scales.

\begin{figure}[ht]
\includegraphics[width=6.3in,angle=0]{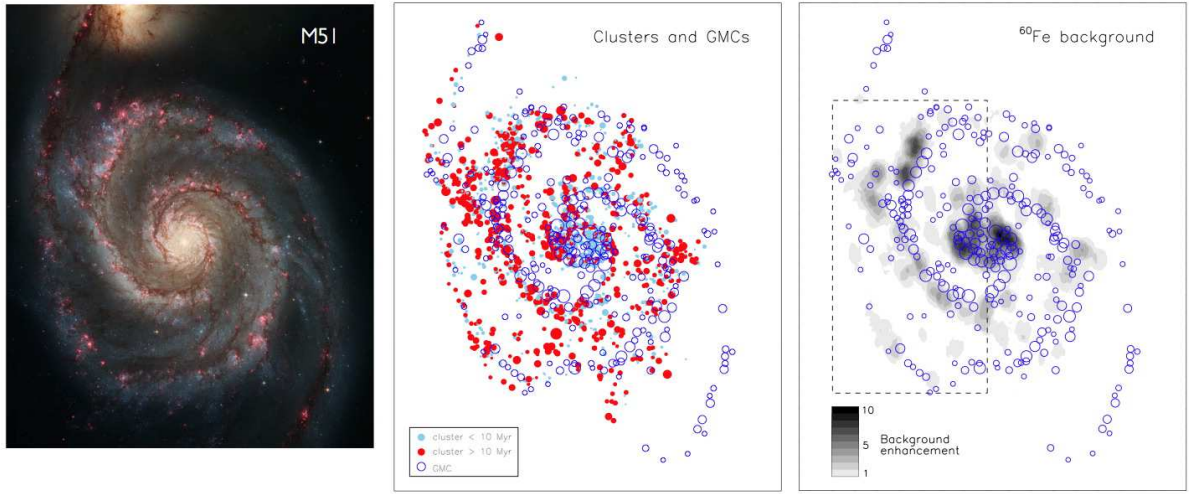}
\caption{Spiral Arm Formation of Enriched (SAFE) clouds.
The left panel shows a Hubble Space Telescope image of the M51 (Whirlpool)
galaxy, a ``grand design'' spiral. The central panel shows the location
of stellar clusters, colour coded by age \cite{2009A&A...494...81S}
and cataloged molecular clouds with symbol size representative
of mass \cite{2009ApJ...700L.132K}.
The clusters overlap in age and are closely correlated with the
location of the molecular clouds in the spiral arms.
The right panel shows the molecular clouds overlaid on a
simple model of the radionuclide background, scaled to the
average within the dashed box.}
\label{fig:m51}
\end{figure}

Due to our location within the Galactic disk, it is hard to
disentangle the projected view of the sky and decipher the
relation between stars and clouds over spiral arm scales.
However, we can detect giant molecular clouds and luminous
clusters in nearby galaxies and make a plausibility test of
this scenario for \fe60\ enrichment.
Figure~\ref{fig:m51} plots the distribution of clouds and
massive stars in the nearly face-on spiral M51
\cite{2009ApJ...700L.132K,2009A&A...494...81S}.
The clusters and clouds follow the spiral arm pattern, of course,
but there is clearly considerable overlap between young clusters,
old clusters, and clouds as might be expected from a feedback
between star and cloud formation. All the clusters shown here
are large with $N_*\simgt 10^3$ and, assuming the same stellar
mass distribution as in the Galaxy, would be expected to contain
several supernova progenitors. Averaged over several Myr, this would
produce a fairly constant flux of ejecta into the surrounding
interstellar medium. To estimate the \fe60\ background,
I calculate the mean mass of supernova ejecta from the luminosity
of each cluster and spread it out as an inverse square law profile
to a radius of 500\,pc, the mean of these {\it superstructures}
\cite{1988ARA&A..26..145T}.
The rightmost panel plots the normalized background, averaged over
the dashed box, enclosing the side of the galaxy where the cluster
identification is complete and including a representative central
region and both spiral arm and interarm regions.
This simple model of the short lived radionuclides background
is highly inhomogeneous
with enhancements of up to a factor of 10 in the centre and some
parts of the arm. Cross-correlating with the cloud catalog shows
that 85\% of the molecular gas mass is enriched above the average
background level and 15\% is enriched by more than a factor of 5.
The factor of 6 enrichment in the \al26\ and \fe60\ levels of the
early solar system relative to the observed $\gamma$-ray average background
is therefore still high but not excessively so, with a percentage
likelihood measured in the double rather than low single digits.
This generalizing of the SPACE scenario to Spiral Arm Formation
of Enriched clouds might be whimsically thought of as locking up
the radionuclides in SAFE clouds.

\section{Summary}
\label{sec:summary}
Stars form in the densest cores of the densest components, molecular clouds,
in the interstellar medium. Most stars form in large clusters
with more than $10^3$ stars, and therefore likely in close
proximity to several massive stars.
The oldest identifiable pieces of meteorites show isotopic
anomalies that indicate the presence of short lived radionuclides
that were incorporated into planetesimals at very early times.
Some may simply have been part of an
approximately steady-state background resulting from a balance
between production in massive stars and decay. Others may have
been produced by nuclear reactions with energetic particles
from an active early Sun. The high abundance of \fe60, however,
shows that the early solar system incorporated fresh
nucleosynthetic products from massive stars.

The winds from evolved stars could impart some short lived radionuclides
into a planetary region but require a highly unlikely chance encounter
between an old star and young star forming region.

The massive stars that likely formed in the same cluster as
the Sun are also highly unlikely to the be the source of this \fe60.
First, the stellar evolutionary timescales are relatively
long compared to disk lifetimes. Second, and more importantly,
protoplanetary disks and star-forming cores are small and
must be very close to a massive star when it explodes as a
supernova to capture the required amount of ejecta. Most objects
would be much further away.

Star formation occurs in many sites within a molecular cloud
and moderate mass clumps on the boundaries of \HII\ regions
can, in principle, capture a \fe60\ mass fraction comparable
to that inferred in the early solar system. The low densities
of these objects, however, imply long collapse times and
there would be substantial decay before incorporation of
the radionuclide into planetesimals.

If we are to explain the \fe60\ levels in our early solar system as a
not-uncommon event, we are forced to consider cloud scales and beyond.
Large clusters produce multiple supernovae which result in an
approximately steady local production of short lived radionuclides
over about 20\,Myr.
Dense gas can result from the shocks at the interfaces of expanding
supernovae ejecta and the enriched material may be rapidly processed into
subsequent generations of stars. The conditions within spiral arms
may be naturally conducive to this and I have presented a simple
plausibility study using catalogs of known clouds and clusters
in M51 to show that spiral arm enhancements of supernova
ejecta may be up to an order of magnitude higher than the average.
Many details remain to be filled in, not the least of which is an
observational demonstration of cloud formation at bubble intersections.

I have focused here exclusively on matching the initial abundance of \fe60.
Supernovae imprint their signatures in other ways including the high
solar system $^{18}$O/$^{17}$O ratio \cite{2010LPI....41.1550Y}.
However the link between the death of a massive star and our origins
occurred, a full accounting needs also to be reconciled, as much as
possible, with the full meteoritic record \cite{1996ApJ...466.1026H}.

The presence of \fe60\ in chondrites presents a major challenge to
theories of star and planet formation. It would be tempting to consider
it simply as a fluke were it not for the possible anthropic implications
\cite{2009Icar..201..821G}.
If the delivery of \fe60, and by extension other short lived
radionuclides including \al26, into our solar system were a rare exception,
the radiogenic baking of the planetesimals in the early solar system would be
much higher and the resulting water content of the planets lower then the norm.
The first ``ocean planet'' has now been discovered \cite{2009Natur.462..891C},
and as the statistics on exoplanet composition build up we
will be able to place the water content of the Earth in context.
Understanding the nature of our birth environment and the consequences thereof,
is an important part of this picture.

\section*{Acknowledgements}
I thank Sasha Krot, Gary Huss, and Eric Gaidos for introducing me
to the puzzle of \fe60\ and the referees for their helpful comments
which improved the manuscript.
My work in this area has benefited greatly from discussions with
Thierry Montmerle, Matthieu Gounelle, Ed Young, and Jeff Hester.
Support for my research comes from the NSF and NASA.

\bibliographystyle{tCPH}
\bibliography{references}

\end{document}